\documentstyle[preprint,aps,prc]{revtex}

\begin{document}
\draft 
\preprint{LBL-39391}
\title{Excess of pions with chiral symmetry restoration}
\author{Chungsik Song and Volker Koch}
\address{Nuclear Science Division, MS 70A-3307\\
         Lawrence Berkeley National Laboratory, Berkeley, CA 94720, USA}
%\date{\today}
\maketitle

\begin{abstract}

We study the effect of the chiral phase transition on pion production 
in hot hadronic matter. 
The phase of restored chiral symmetry is characterized by the appearance of the
scalar $\sigma$-meson as a chiral partner of the pion as well as by the
degeneracy of the vector and axial-vector mesons. 
We find rapid thermal and chemical equilibration of these degrees of freedom in
the symmetric phase. Provided that the chiral transition temperature
is not considerably high, the presence of a
chirally symmetric phase will result in $\sim 1.6$ times more thermal 
pions in the final state. 
\end{abstract}
\bigskip
\pacs {}
%PACS Numbers : 25.75.+r,25.80.Ek,12.38.Mh,11.30.Rd}

\section {}
  
Chiral symmetry, a symmetry of quantum chromodynamics (QCD) 
in the limit of massless quarks, is spontaneously
broken in the ground state of QCD as indicated 
by the small mass of the pion \cite{georgi}. 
At high temperatures chiral symmetry is expected to be restored, 
as shown by numerical simulations in lattice QCD 
\cite{lattice} as well as  model calculations \cite{model}.
However, some intriguing questions still remain on how chiral
symmetry is actually restored in hot hadronic matter and what the 
signatures of the restored phase are.
High energy nucleus-nucleus collisions 
offer a unique opportunity to explore the properties of 
hadronic matter at high temperatures and densities and to address these
questions in experiment.

One way to study the signatures of the chiral phase transition in hot
hadronic matter is by comparing the properties of the hadronic system in the 
broken phase with those  in the symmetric phase. 
One of the interesting features  of the symmetry restored phase is the 
appearance of the scalar $\sigma$-meson which forms a chiral  multiplet with
the pions.  The $\sigma$-meson corresponds to an amplitude fluctuation of 
the order parameter, the quark condensate,  while the 
pions are related to a phase fluctuation. At low temperatures where chiral
symmetry is spontaneously broken, the $\sigma$-meson has very
large width due to the strong decay channel into two pions. 
For that reason it is very difficult to observe  any resonance peak in 
the scalar channel of the $\pi-\pi$ phase shift.
In a 
recent analysis of the $I=0$ $S$-wave $\pi\pi$ phase shift $\delta^0_0$, the
existence of a $\sigma$ resonance has been inferred with $m_\sigma=535\sim 650$
MeV \cite{s_exp}. 

On the other hand, as the quark condensate drops with increasing 
temperature, the mass difference of the $\sigma$-meson and pion becomes 
small. As a result, the decay width of scalar meson decreases because the
phase space available for the outgoing pions is reduced.
Close to the phase transition temperature $T_\chi$, where the $\sigma$-mass has
dropped below that of two pions, the decay channel closes and the $\sigma$
becomes an elementary excitation.  
The observation of a narrow width scalar meson has
been suggested as a direct signature of chiral symmetry restoration
\cite{sigma}, but since the $\sigma$ does not couple to any penetrating 
probes such as
photons and dileptons this is difficult to observe in experiment.
 
The purpose of this paper is to study the effect of the $\sigma$-meson
on the pion production in high energy nucleus-nucleus collisions. 
The pion production will depend on
the thermodynamic equilibration conditions of a system during the evolution
from hadronization to freeze-out. 
We will show that the chemical equilibrium conditions in the chirally symmetric
phase are quite different from those in the broken phase 
mainly due to the appearance of the light $\sigma$  meson and the drop of the
chiral condensate. This will have some interesting consequence for the final
state pion abundance.

\section{}

Let us consider the $SU(2)_L\times SU(2)_R$ linear sigma model in order 
to describe the hadronic system close to the chiral phase transition 
\cite{linear}. All the arguments presented below will only depend on the
smallness of the chiral condensate and the $\sigma$-mass. Therefore, we will
consider the system only in the hadronic phase -- possibly close to the chiral
phase transition temperature-- and ignore effects from deconfinement.
We furthermore assume that the axial $U_A(1)$ symmetry is not restored at the
temperatures under considerations, which is supported by recent lattice
calculations \cite{Laer}. 
The linear sigma model includes isotriplet pseudoscalars ($\pi^a$), the pions, 
and the isosinglet scalar ($\sigma$). 
They form a $({1\over2},{1\over2})$ representation of 
$SU(2)_L\times SU(2)_R$, and are grouped into 
\begin{equation}
\Phi={1\over2}
(\sigma{\bf 1} +i\mbox{\boldmath$\tau$}\cdot\mbox{\boldmath$\pi$}),
\end{equation}
where ${\bf 1}$ is the $2\times 2$ unit matrix and 
the $\mbox{\boldmath$\tau$}$'s are the Pauli spin matrices. 

We furthermore include vector $(\mbox{\boldmath $\rho$}^\mu)$ 
and axial-vector $(\mbox{\boldmath $a$}_1^\mu)$ mesons as gauge fields 
and the total Lagrangian is
given by  \cite{lvector,pisarski}
\begin{eqnarray}
{\cal L}_{eff}&=& {\rm Tr}D_\mu\Phi^* D^\mu\Phi
                 + \mu^2 {\rm  Tr}\,(\Phi^*\Phi)
                 - \lambda\, {\rm  Tr}\,(\Phi^*\Phi)^2
\nonumber\\[2pt]
               &&-{1\over4}{\rm Tr}(F_{L\,\mu\nu}F_L^{\mu\nu}
                                   +F_{R\,\mu\nu}F_R^{\mu\nu})
\nonumber\\[2pt]
               &&+{1\over2}m_0^2\,{\rm Tr}\,(A_{L\,\mu}A_L^\mu+A_{R\,\mu}A_R^\mu)
\nonumber\\[2pt]
               &&-h{\rm Tr}(\Phi),
\end{eqnarray}               
where 
$A_{L}^\mu=(\mbox{\boldmath $\rho$}^\mu+\mbox{\boldmath $a$}_1^\mu)
\cdot\mbox{\boldmath$\tau$}/2$,
$A_{R}^\mu=(\mbox{\boldmath $\rho$}^\mu-\mbox{\boldmath $a$}_1^\mu) 
\cdot\mbox{\boldmath$\tau$}/2$ and
\begin{equation}
D^\mu\phi=\partial^\mu\phi+ig A_L^\mu\phi-ig \phi A_R^\mu.
\end{equation}
The last term is responsible for the explicit breaking of chiral symmetry due
to the small but finite mass  of the light quarks.  
$\mu$, $\lambda$, $g$, and $h$ are constants which are inferred by fitting
the pion decay constant, the vector meson decay width, the sigma meson mass 
as well as the mass of the pion.

In the broken phase, at low temperatures, the scalar field has a 
non-vanishing vacuum expectation value, 
$\langle\sigma\rangle\equiv\sigma_0
=(\mu^2/\lambda)^{1/2}$, 
which plays the role of the
order parameter of the chiral phase transition. After redefining the field 
$\sigma\to\sigma+\sigma_0$, one obtains massless Goldstone boson, 
$m_\pi=0$, and a massive scalar meson, $m_\sigma=\sqrt{2\lambda}\sigma_0$ 
in the chiral limit, $h=0$. The degeneracy between vector and axial-vector 
meson
masses is also broken due to the finite expectation value of the
$\sigma$-field; 
$m_\rho^2=m_0^2$ and $m_{a_1}^2=m_0^2+g^2\sigma_0^2$.

Because of the spontaneous breaking of chiral symmetry, the derivative of the 
pion field contributes to the axial current and, therefore, the pion and axial
vector field mix. 
To obtain physical fields we redefine the axial gauge fields as 
\begin{equation}
\mbox{\boldmath $a$}_1^\mu\to \mbox{\boldmath $a$}_1^\mu-{\chi\over Z}
\biggl( \partial^\mu\mbox{\boldmath$\pi$}
      -ig[\mbox{\boldmath$\rho$}^\mu,\mbox{\boldmath$\pi$}]\biggr),
\end{equation}
where $\chi=g\sigma_0/(g^2\sigma_0^2+m_0^2)$. $Z$ is the pion wave function
renormalization constant and given by 
\begin{equation}
Z^2=1-{g^2\sigma_0^2\over (g^2\sigma_0^2+m_0^2)}.
\end{equation}

In the broken phase we have a strong decay channel for the
scalar meson into two pions, namely 
\begin{equation}
{\cal L}_{\sigma\pi\pi}=
-{\lambda\sigma_0\over Z^2}
                          \sigma\mbox{\boldmath $\pi$}^2
+{g\chi\over Z^2}\left[\partial_\mu\sigma\mbox{\boldmath $\pi$}\cdot
   \partial^\mu \mbox{\boldmath $\pi$}
-(1-g\sigma_0\chi)\sigma\partial^\mu \mbox{\boldmath $\pi$}\cdot
   \partial^\mu \mbox{\boldmath $\pi$}\right]
\end{equation}
where the parameter $\lambda$ is determined by the sigma meson mass. 
It assumes a value of 21.4 (7.62) for  $m_\sigma\sim 1$ GeV (0.6 GeV). 
With this coupling the 
resulting width is $\Gamma_{\sigma\pi\pi}\simeq 800\,(600) \>\rm MeV$.

In the chirally symmetric phase, the expectation value of the $\sigma$-field
vanishes, $\sigma_0=0$. In this case, the $\sigma$-meson and
the pions as well as the vector- and axial-vector fields 
form a degenerate chiral multiplet, respectively.
As a consequence the decay of the $\sigma$-meson into two pions is
kinematically forbidden. Also the vector and axial-vector current-current 
correlators 
will be identical, a property which persists at finite temperature 
\cite{pisarski}.

Moreover, the coupling constants among the mesons will be modified 
in  the symmetric phase since they depend on the strength of the scalar 
condensate. 
In the chirally symmetric phase we have 
\begin{eqnarray}
{\cal L}_{\pi\pi\rho} &\rightarrow&  g\mbox{\boldmath $\rho$}^\mu\cdot 
                              (\mbox{\boldmath $\pi$}\times
                              \partial_\mu\mbox{\boldmath $\pi$}),
\nonumber\\[2pt]
{\cal L}_{\pi\pi\sigma} &\rightarrow & 0,
\nonumber\\[2pt]
{\cal L}_{\pi\rho a_1}  &\rightarrow & 0,
\nonumber\\[2pt]
{\cal L}_{\pi\sigma a_1} &\rightarrow& -g\mbox{\boldmath $a$}_1^\mu\cdot 
                              (\sigma\partial_\mu\mbox{\boldmath $\pi$} 
                              -\partial_\mu\sigma\mbox{\boldmath $\pi$}).
\label{coupl}
\end{eqnarray}               
We can see that  the coupling of 
$\sigma$ meson to two pions vanishes, that is, 
the decay channel for scalar mesons into two pions is completely closed.
Also the $\pi-\rho-a_1$ coupling vanishes in the chirally symmetric phase.
Only vector and axial vector mesons decay into two pions, and pion and
scalar meson, respectively.
Since the couplings are proportional to the scalar
condensate, they become already negligible once one gets 
close enough to $T_\chi$
{\em without} actually restoring chiral symmetry. 

The dependence of the coupling constants on
the chiral condensate leads to interesting consequences related to the chiral
phase transition. For example the thermal
photon production in hadronic matter would be affected considerably. The
dominating production channel is believed to be the reaction 
$\pi\rho\to a_1\to\pi\gamma$ \cite{song}
which is due to the interaction 
\begin{equation}
{\cal L}_{a_1\rho\pi}=-{g\over Z}\chi\biggl[
                       (\partial_\mu \mbox{\boldmath $\rho$}_\nu
                       -\partial_\nu \mbox{\boldmath $\rho$}_\mu)\cdot
                        \partial^\mu\mbox{\boldmath $\pi$}\times 
                                    \mbox{\boldmath $a$}_1^\nu
                      +(\partial_\mu \mbox{\boldmath $a$}_{1\,\nu}
                       -\partial_\nu \mbox{\boldmath $a$}_{1\,\mu})\cdot
                        \partial^\mu\mbox{\boldmath $\rho$}^\nu\times
                                    \mbox{\boldmath $\pi$}\biggr]
\end{equation}
assuming vector meson dominance. 
This reaction is shut off in the symmetric phase since
$Z\to 1$ and $\chi\to 0$ as the condensate ($\sigma_0$) vanishes, implying
a suppression of the photon production in the chirally symmetric phase \cite
{song_photon}.

\section{}

In this section we study the consequences of the chiral phase transition 
related to the chemical equilibration of pions. 
To this end we will calculate the chemical equilibration times in the broken as
well as in the chirally symmetric phase. Thermodynamic equilibration 
is driven by the multiple collisions among the particles in the system.
When the collision rate is fast compared to the lifetime of the hadronic 
system thermodynamic equilibrium can be reached.
In both phases, broken as well as symmetric, thermal, i.e. kinetic 
equilibration of pions  
is governed by elastic two-body collisions, $\pi\pi\rightleftharpoons\pi\pi$.
At temperatures $T >  100 \, \rm MeV$, this reaction is dominated by the strong
p-wave isovector ($\rho$) channel. Since the coupling to the $\rho$-meson is
not changed in the symmetric phase (see eq. (\ref{coupl}) ), 
this reaction will be equally strong in both phases. In addition, in the
symmetric phase, the s-wave scattering will be considerably stronger than in
the broken phase where these reactions are small due to the Goldstone nature
of the pions.   
It has been shown that already in the broken phase 
elastic collisions are frequent enough  
for pions to maintain thermal equilibrium even at low temperatures 
\cite{thermal}. Since there are the additional s-wave interactions in the
symmetric phase, it is safe to assume that kinetic equilibrium will be
maintained in both phases to a very good approximation.

Chemical
equilibration, on the other hand involves  inelastic processes 
which only contribute in the next to leading order of the low energy expansion.
As we shall see in the following, the disappearance of some of the couplings
(\ref{coupl}) in
the symmetric phase will lead to quite different chemical equilibration
conditions from these in the broken phase.

\subsection{In symmetric phase}

{\it (i)} $\pi\pi\rightleftharpoons\pi\pi\pi\pi$

\begin{figure}
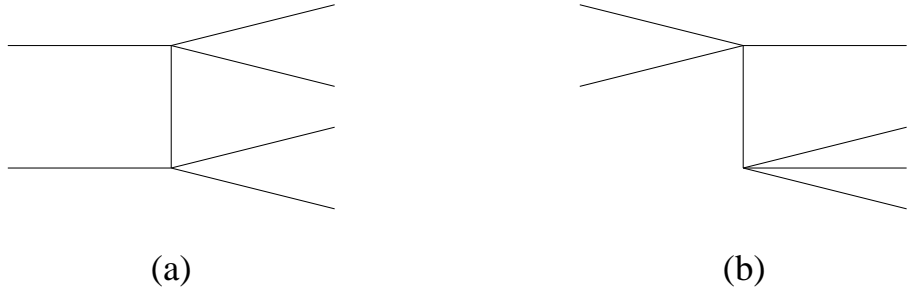

\caption{Diagrams for inelastic scattering reactions, 
$\pi+\pi\leftrightarrow\pi+\pi+\pi+\pi$. 
There are diagrams not shown explicitly, which can be obtained by 
interchanging the final state.
\label{feyn_pppp}}
\end{figure}

\noindent
Chemical equilibrium is caused by the various inelastic, particle-number
changing reaction.
Contrary to the broken phase, we have  strong s-wave
$\pi\pi\rightleftharpoons\pi\pi\pi\pi$ interactions in the symmetric phase
as shown in Fig.~\ref{feyn_pppp}.
This reaction leads to an {\em absolute} chemical
equilibrium of pions in medium which is defined by $\mu_\pi=0$ 
\footnote{For a definition of the meaning of a pion chemical equilibrium 
in the context of heavy ion collisions see ref. \cite{chemical}.}. 
The chemical relaxation time is given by \cite{chemical}
\begin{equation}
{1\over\tau_{ch}}={4\over n_\pi^0} I_0(\pi\pi\leftrightarrow\pi\pi\pi\pi; T)
\end{equation}
where $n_\pi^0$ is the pion density at equilibrium and  
\begin{eqnarray}
I_0(\pi\pi\leftrightarrow\pi\pi\pi\pi; T)&=& {\cal S} \,
\int{d^3p_1\over (2\pi)^3 2E_1}\int{d^3p_2\over (2\pi)^3 2E_2}
\cdots\int{d^3p_6 \over (2\pi)^3 2E_6}
\nonumber\\[4pt]
&&
\times \,\, (2\pi)^4\delta^{(4)}(p_1 + p_2 - p_{3} - \cdots-p_6)
\nonumber\\[4pt] 
&&
\times \sum_{1,2,\ldots,5,6}\vert{\cal M}
(\pi_1\pi_2\leftrightarrow\pi_3\cdots\pi_6 )
\vert^2 e^{-(E_1+E_2)/T}.
\label{izero}
\end{eqnarray}
Here $\cal S$ indicates the symmetric factors. Here, and throughout this paper,
we use the Boltzmann limit of the Bose-Einstein quantum statistics.

The calculation has been done ignoring  
the interference terms among the six different diagrams \cite{japan}. 
At a temperature  $T = 180 \, \rm MeV$ we find  for the chemical relaxation
time $\tau_{ch}= 2.8 \, \rm fm$ 
corresponding to the couplings $\lambda =21.4 $ ($m_{\sigma} = 1 \, \rm GeV$).
Since the cross section is proportional to 
$\lambda^4$ the results strongly depend on the values of $\lambda$.
From the result we can see that it might be possible for pions to 
achieve an {\em absolute}
chemical equilibrium with $\mu_\pi=0$ near the symmetry restored phase for 
large values of $\lambda$. However, the value, of the coupling $\lambda$
depends on the assumed value of the mass of the $\sigma$ meson, which is not
well determined. Furthermore, higher order thermal effect may
renormalize the value of $\lambda$. Therefore, we cannot make any definite
statements about the possibility of an absolute chemical equilibrium in the
symmetric phase. 

There are also inelastic reactions involving vector and axial vector
mesons such as $\pi\pi\rightleftharpoons a_1 a_1$ or $\pi\pi\rightleftharpoons
\rho\rho$. These reactions, which would drive the system to an 
absolute chemical equilibrium ($\mu_\pi\to 0$), however, turn out to be as 
slow as in the broken phase, $\tau_{ch}\sim 7 \rm\>fm/c$ 
at $T=180\rm\> MeV$ \cite{chemical}. 

\bigskip

{\it (ii)} $\pi\pi\rightleftharpoons\sigma\sigma$

\begin{figure}
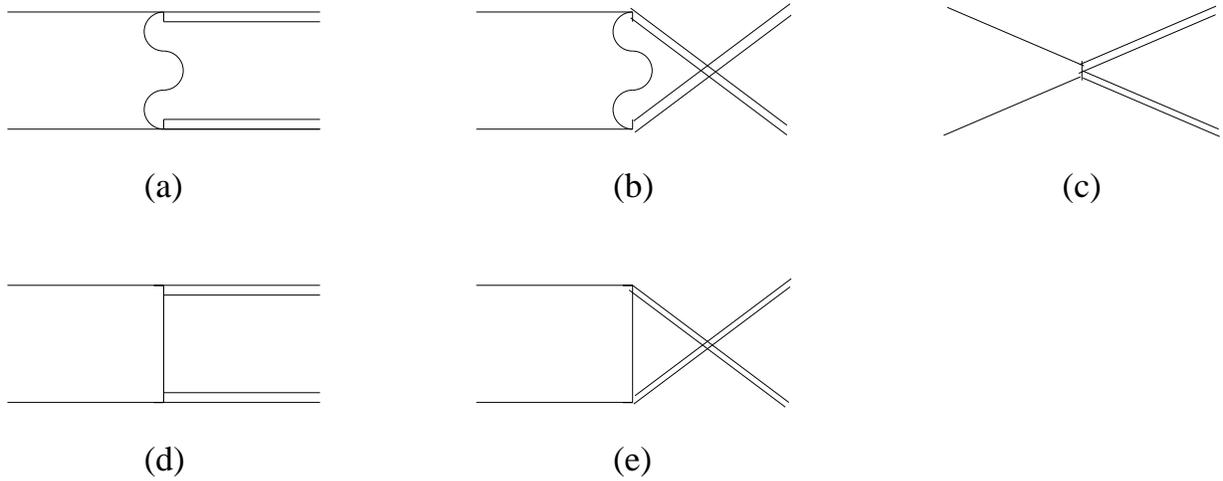

\caption{Diagrams for inelastic scattering reactions, 
$\pi+\pi\to\sigma+\sigma$. The solid line, double solid line and wave
indicate pion, sigma meson and axial vector meson, respectively.}
\label{feyn_ppss}
\end{figure}

\noindent
Next there is the reactions $\pi\pi\rightleftharpoons\sigma\sigma$ 
(Fig.~\ref{feyn_ppss}). Processes involving the exchange of an  
axial vector meson (Fig.~\ref{feyn_ppss}.a and \ref{feyn_ppss}.b) as well the 
the direct coupling of two pions to two sigma mesons 
(Fig.~\ref{feyn_ppss}.c) contribute. The latter is given by 
\begin{equation}
{\cal L}_{\pi\pi\sigma\sigma}=
{\lambda\over 2Z^2}\sigma^2\mbox{\boldmath $\pi$}^2
+{g^2\chi^2\over 2Z^2}\sigma^2(\partial_\mu\mbox{\boldmath $\pi$})^2.
\end{equation}
These reactions lead to  {\em relative} chemical equilibrium for which
$\mu_\pi=\mu_\sigma$ but $\mu_\pi$ need not to be zero.

\begin{figure}
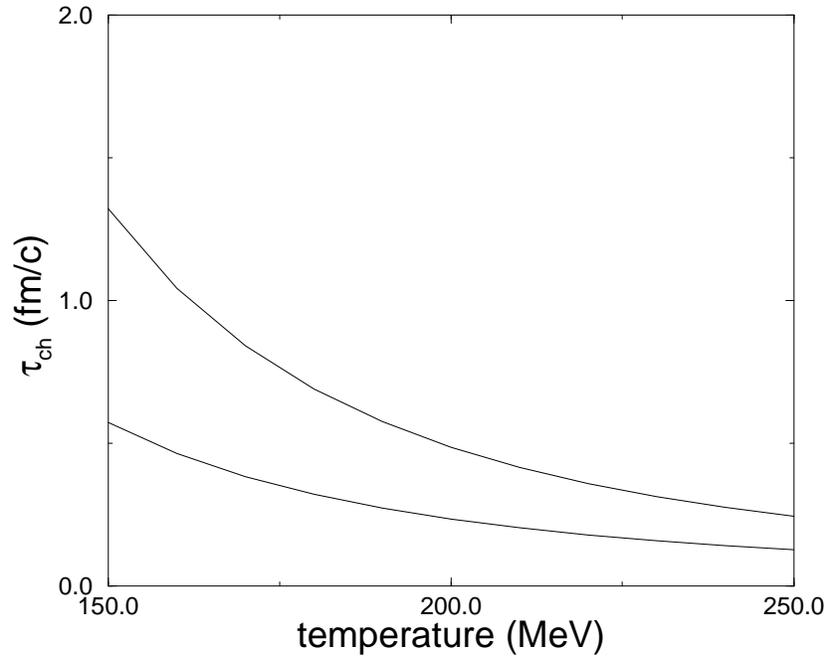

\caption{Chemical relaxation time of pions at finite temperature. 
The results are 
obtained in the chirally symmetric phase with $m_\pi=m_\sigma=140\>\rm MeV$ 
and $m_\rho=m_{a_1}=770 \>\rm MeV$. 
Two different values of $\lambda =21.4 \>\rm (lower), \>7.62\>\rm (upper)$ 
are used.  
\label{t_symmetry1}}
\end{figure}

The {\em relative} chemical relaxation time is given by \cite{chemical}
\begin{equation}
{1\over\tau_{ch}}=4I_0(\pi\pi\rightleftharpoons\sigma\sigma; T)
\left({1\over 3 n_\pi^0} + {1\over n_\sigma^0}\right)
\end{equation}
where $I_0(\pi\pi\rightleftharpoons\sigma\sigma; T)$ is given by the similar
form as that of Eq.~\ref{izero} for the reaction 
$\pi\pi\rightleftharpoons\sigma\sigma$.
In the calculation of the scattering amplitude 
we introduce a form factor for the vertex 
involving the exchange of axial vector meson, given by
\begin{equation}
F={\Lambda^2-m_{a_1}\over \Lambda^2-t}
\end{equation}
where we use the values of $\Lambda=1.7 \,\rm GeV$.
The results for the relaxation time are shown in Fig.~\ref{t_symmetry1} 
as a function of temperature assuming that the degenerate masses of the 
scalar and vector mesons are $m_\pi=m_\sigma=140\,\rm MeV$ 
and $m_\rho=m_{a_1}=770\,{\rm MeV}$. 
Two different values of the $\lambda=$21.4 and 7.62
are considered corresponding to 
$m_\sigma=$1 GeV and 0.6 GeV respectively. 
We can see that the relaxation time is about 0.3 $\sim$ 0.6 fm at $T=180$ MeV 
which is much shorter than the effective size of the hot matter, 
$R=2\sim 3 \,\rm fm$ \cite{chemical}.
This implies that 
pions will reach {\em relative} chemical equilibrium with 
respect to the reaction
$\pi\pi\rightleftharpoons\sigma\sigma$ with $\mu_\pi=\mu_\sigma$ in the 
chirally symmetric phase. 

\begin{figure}
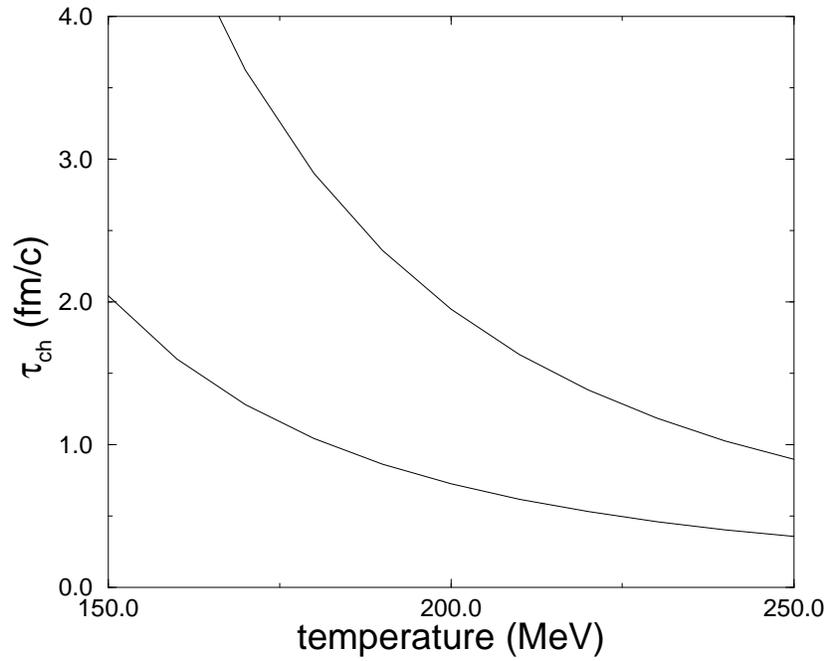

\caption{Chemical relaxation time of pions at finite temperature. 
The results are 
obtained in the chirally symmetric phase with thermal effects on the meson
mass, $m_\pi=m_\sigma=220\>\rm MeV$ and $m_\rho=m_{a_1}=980 \>\rm MeV$. 
Two different values of $\lambda =21.4\>\rm (lower),\> 7.62\>\rm (upper)$
are used.  
\label{t_symmetry2}}
\end{figure}

In order to take into account thermal corrections to the effective mass of 
the mesons \cite{pisarski} 
we also show the chemical relaxation times for  
$m_\pi=m_\sigma=220\, \rm MeV$ and $m_\rho=m_{a_1}=980 \,\rm MeV$
in Fig. \ref{t_symmetry2}.
In this case the reaction times are considerably slower but it is still
possible to reach  chemical equilibrium near the critical temperature, which
is $T_\chi\approx 220 \,\rm MeV$ in this model.

In addition, there are the decays and formation of the vector meson, which are
according to eq. (\ref{coupl}), $\rho\rightleftharpoons\pi\pi$, 
$a_1\rightleftharpoons\pi\sigma$. Notice, that in the symmetric phase 
the $a_1$ could not decay into $\pi \rho$. These decays are fast and will lead
to a relative chemical equilibrium characterized by   $\mu_\rho=2\mu_\pi$ and 
$\mu_{a_1}=\mu_\pi+\mu_\sigma$. Taking into account the relative equilibrium
between pions and the $\sigma$ meson, we arrive at the following chemical
equilibration conditions in the symmetric phase:
$\mu_\pi = \mu_\sigma$, $\mu_\rho=2\mu_\pi$,  $\mu_{a_1}=2 \mu_\pi$

\subsection{In broken phase}

Recently \cite{chemical}, we have shown that pions can easily reach a 
{\em relative}
chemical equilibrium  with the $\rho$-mesons while the reaction times 
leading to 
a {\em absolute} chemical
equilibrium, i.e. $\mu_\pi = 0$  are too long to be effective at temperatures
$T<160 \, \rm MeV$.
For the reactions involving scalar mesons we expect a
{\em relative} chemical equilibrium of 
pions with the $\sigma$-mesons through the 
reaction $\sigma\rightleftharpoons\pi\pi$,
since the decay width of $\sigma\to\pi\pi$ is very large.
Assuming {\em relative} chemical 
equilibrium between pions and scalar mesons
the reaction  $\pi\pi\rightleftharpoons\sigma\sigma$ 
is effectively identical to the reaction 
$\pi\pi\rightleftharpoons\pi\pi\pi\pi$ and, thus, leads to  
{\em absolute} chemical equilibrium. 
The same reaction channel has been considered in the symmetric phase and shown
to be very fast. In broken phase  the dominant contribution comes 
from the pion
exchange diagrams as shown in Fig.~(\ref{feyn_ppss}.d) and
(\ref{feyn_ppss}.e). 
However, the leading contribution 
proportional to $\lambda^4$ is canceled by the diagrams with four pions as in
Fig.~\ref{feyn_pppp} and
the result is very similar to that obtained in chiral perturbation theory
\cite{goity},  which is  $\tau_{ch} \simeq 200 \, \rm fm$. 

In the broken phase at a temperature of about $T\sim 160$ MeV, therefore, 
a {\em relative} chemical equilibrium is maintained and it is characterized by
$\mu_{\sigma} = 2 \mu_\pi$, $\mu_{\rho} = 2 \mu_\pi$, $\mu_{a_1} = 3 \mu_\pi$.
This is different from the conditions in the symmetric phase.

\section{}

Finally, we study observable consequence of the different chemical 
relaxation conditions in the broken and symmetric phase.
In the symmetric phase, $m_\pi = m_\sigma$ and $\mu_\pi = \mu_\sigma=\mu$. 
Therefore,
we will have as many $\sigma$ mesons as one third of 
pions,
\begin{equation}
N_\sigma=\frac{1}{3}N_\pi=\frac{1}{3}
\biggl(N_{\pi^+}(T,\mu)+N_{\pi^-}(T,\mu)+N_{\pi^0}(T,\mu)\biggr)
\end{equation}
Similar for the vector mesons we have 
$m_{a_1} = m_\rho$ and $\mu_{a_1} = 2 \mu_\pi =
\mu_\rho$. 
Thus there are
as many  $a_1$ mesons as $\rho$ mesons in the symmetric phase.

In the broken phase, on the other hand, we have 
$\mu_\sigma=2\mu_\pi=2\mu$ and $m_\sigma \gg m_\pi$ and, therefore, 
\begin{equation}
N_\sigma\sim e^{-m_\sigma/T}\ll N_\pi,
\end{equation}
as long as $\mu_\pi \ll m_\sigma$.
For the vector and axial vector mesons the chemical equilibrium conditions are 
$\mu_\rho=2\mu_\pi$, $\mu_{a_1}=\mu_\pi+\mu_\rho=\mu_\pi+\mu_\sigma=3\mu_\pi$. 

Given the same chemical potential in both phases, there will be  
more sigma mesons and axial vector mesons in chirally symmetric phase
than in the broken phase simply due to the mass difference.
This abundance can be observed in the final state pion number, 
if all inelastic channels,   except
the decay process, $\sigma\rightleftharpoons\pi\pi$,
$a_1\rightleftharpoons\pi\rho$, are inactive as the system passes through 
the broken phase.
For a simple estimate 
we assume a chirally symmetric phase at the early stage of
the hot matter with $\mu_\pi=0$. 
Since all resonances decay into pions the observed 
pion number will be 
\begin{eqnarray}
\bar N_\pi(T_f) &=& N_\pi(T_h)+2N_\sigma(T_h)+\cdots\cr 
        &\approx& {5\over3} N_\pi(T_h)+\cdots.
\end{eqnarray}               
where we use the fact that $N_\sigma={1\over3}N_\pi$. Here we do not show 
the contribution from $a_1$ mesons explicitly since $N_{a_1}\ll N_\pi$. 
$T_h$ is either the hadronization temperature or initial temperature of the
hot hadronic matter and $T_f$ is the freeze-out temperature.
Alternatively, let us assume that there is no chirally symmetric phase at the
initial stage of hot hadronic matter. 
Since the number of sigma mesons is almost negligible
the observed pion number will be 
\begin{equation}
\bar N_\pi(T_f) \approx N_\pi(T_h)+\cdots.
\end{equation}               
When there is a chirally symmetric phase initially, 
thus, we have about 1.6 times more thermal pions. 

However, this result will be modified when there are pion number
changing reactions in the system at low temperatures. 
In our previous work \cite{chemical}, we have shown that the reaction rates
for the pion number changing processes will depend on the pion chemical
potential as well as temperature. 
If we assume a chirally symmetric phase
initially, which then turns into a phase of broken symmetry where the mass of
the $\sigma$-meson is large, the number of
$\sigma$-mesons will be oversaturated. 
These $\sigma$-mesons will rapidly decay leading to a finite chemical potential
for the pions. Similarly, since the mass of the $a_1$ mesons increases as
chiral symmetry gets broken, they also will be oversaturated leading to an
additional increase of the pion chemical potential. 

To estimate the induced pion chemical potential we assume
that the chiral condensate instantly build up at $T=T_\chi$ and $\mu_\pi=0$
initially.  
Since the effective number of pions should be constant at temperatures both
slightly above ($T=T_\chi^+$) and below ($T=T_\chi^-$) the critical
temperature, we have 
\begin{equation}
\bar N_\pi(T_\chi^+,\mu_\pi=0)=\bar N_\pi(T_\chi^-,\mu_\pi)
\end{equation}
where $\bar N_\pi(T)=N_\pi+2N_\sigma+2N_\rho+3N_{a_1}$.  
From this we have $\mu_\pi=55\sim 66$ MeV at $T_\chi=160\sim 200$ MeV,
respectively.
If $T_\chi\sim 160$ MeV the chemical relaxation time of pions, with
$\mu_\pi=55$ MeV, will be about 3 fm and is comparable to the 
effective size of the hot hadronic system \cite{chemical}. 
If the chiral phase transition occurs at higher temperature than $T=160$ MeV
pion number changing processes will be active and the number of pions will be
reduced. Thus we expect less pions than that obtained when we 
completely neglect the pion number changing
processes as long as $T_\chi>160\>\rm MeV$.

\section{}

In summary, we have studied the chemical equilibration conditions of pions
in hot hadronic matter  assuming chiral symmetry restoration. 
The chirally symmetric phase is described with 
light scalar mesons which are degenerate with pions, and with 
degenerate vector and axial-vector mesons.
In the symmetric phase a  {\em relative} 
chemical equilibrium among the particles will be established rapidly, leading
to $\mu_\sigma = \mu_\pi $ and $\mu_\rho = 2 \mu_\pi = \mu_{a_1}$.
The number of scalar mesons is then given by one third of 
the total number of pions at the given temperature. 
This is quite different from the situation encountered in the  broken
phase, where $\mu_\sigma = 2 \mu_\pi$ and $\mu_{a_1} = 3 \mu_\pi$.

This difference in chemical equilibration conditions of pions 
might lead to the excess of pions at freeze-out. 
As temperature decreases and the symmetry is broken 
scalar mesons become heavier and decay into two pions. Also $a_1$ mesons decay
into three pions.  
The number of observed pions will be given by the number of pions plus
contributions from the resonance decay.
When we include the scalar and $a_1$ meson contributions to observed pions, 
we have $\sim 1.6$ times more pions 
compared to the case when there is no chiral phase transition.
This, however, can only be observed, if the chiral transition temperature
is not too high, $T_{\chi} \leq 160$ MeV. Otherwise, pion number changing
processes in the broken phase will absorb the excess obtained from the chiral
phase transition.

Finally, let us conclude by pointing out that an analysis of the particle
abundances measured at SPS-energy heavy ion collisions found an excess of pions
by a factor of 1.6 over the expected thermal value \cite{gadiz}. 
However, a different analysis based on the same data, does not find such an
excess of pions \cite{stachel}.  

\bigskip
This work supported by the Director, Office of Energy Research, Office of High
Energy and Nuclear Physics, Division of Nuclear Physics, Division of Nuclear
Sciences, of the U. S. Department of Energy under Contract No. 
DE-AC03-76SF00098.

%%%%%%%%%%%%%%%%%%%%%%%%%%%%%%%%%%%%%%%%%%%%%%%%%%%%%%%%%%%%%%
%%%%%%%%%%%%%%%%%%%%%%%%%%%%%%%%%%%%%%%%%%%%%%%%%%%%%%%%%%%%%%

%%%%%%%%%%%%%%%%%%%%%%%%%%%%%%%%%%%%%%%%%%%%%%%%%%%%%%%%%%%%%%
%%%%%%%%%%%%%%%%%%%%%%%%%%%%%%%%%%%%%%%%%%%%%%%%%%%%%%%%%%%%%%

%%%%%%%%%%%%%%%%%%%%%%%%%%%%%%%%%%%%%%%%%%%%%%%%%%%%%%%%%%%%%%
%%%%%%%%%%%%%%%%%%%%%%%%%%%%%%%%%%%%%%%%%%%%%%%%%%%%%%%%%%%%%%

%%%%%%%%%%%%%%%%%%%%%%%%%%%%%%%%%%%%%%%%%%%%%%%%%%%%%%%%%%%%%%
%%%%%%%%%%%%%%%%%%%%%%%%%%%%%%%%%%%%%%%%%%%%%%%%%%%%%%%%%%%%%%


\begin{references}
%\begin{thebibliography}{20}

\bibitem{georgi} H. Georgi, {\it Weak Interactions and Modern Particle Theory}
                 (The Benjamin/Cummings Publishing Company, Inc., 
                  Menlo Park, California, 1984).

\bibitem{lattice} C. de Tar, in {\em Quark Gluon Plasma 2}, R. Hwa, editor,
                  World Scientific, Singapore, 1995.

\bibitem{model} T. Hatsuda and T. Kunihiro, Phys. Rev. Lett. 55 (1985) 158.\\
                J. Gasser and H. Leutwyler, Phys. Lett. B 184 (1987) 83.

\bibitem{s_exp} S. Ishida et. al., Prog. Theor. Phys. 95 (1996) 745.

\bibitem{sigma} T. Hatsuda and T. Kunihiro, 
                Prog. Theor. Phys.  Suppl. 91 (1987) 284.\\
                H. A. Weldon, Phys. Lett. B 274 (1992) 133.\\
                R. Pisarski, Phys. Rev. Lett. 76 (1996) 3084.

\bibitem{linear} Benjamin W. Lee, {\it Chiral Dynamics} 
                 (Gordon and Breach Science Publishers, 
                  New York, New York, 1972);
                 Volker Koch, LBL-39463, to appear in 
                 Int. Jou. Mod. Phys. E (June 1997).

\bibitem{Laer} E. Laermann, Nucl. Phys. A 610 (1996) 1c.

\bibitem{lvector} P. Ko and S. Rudaz, Phys. Rev. D 50 (1994) 6877.

\bibitem{pisarski} R. Pisarski, Phys. Rev. D 52 (1996) 3773.

\bibitem{song} L. Xiong, E. Shuryak and G. E. Brown, 
               Phys. Rev. D 46 (1992) 3798.\\
               Chungsik Song, Phys. Rev. C 47 (1993) 2861.

\bibitem{song_photon} Chungsik Song (in preparation).

\bibitem{thermal} J. L. Goity and H. Leutwyler, 
                  Phys. Lett. B 228 (1989) 517.\\
                  Chungsik Song, Phys. Rev. D 48 (1994) 1556.\\
                  K. Haglin and S. Pratt, Phys. Lett. B 328 (1994) 255.

\bibitem{japan} Mitsuru Ishii, Yukawa Institute preprint, (hep-ph/9608265).

\bibitem{chemical} Chungsik Song and Volker Koch, 
                   LBL-38363 (1996), submitted to Phys. Rev C.

\bibitem{goity} J. L. Goity, Phys. Lett. B 319 (1993) 401.

\bibitem{gadiz} M. Gazdzicki for the NA35 Collaboration, 
                Nucl. Phys. A 590 (1995) 215c.

\bibitem{stachel} P. Braun-Munzinger, J. Stachel, J. Wessels and N. Xu,
                  Phys. Lett. B 365 (1996) 1.


%\end{thebibliography}

\end{references}
\end{document}